\documentclass[prl,twocolumn,showpacs,floatfix]{revtex4}

\usepackage{graphicx}
\usepackage{amsmath}

\begin{document}
\title{Mimicking Time Evolution within a Quantum Ground State: Ground-State Quantum Computation, Cloning, and Teleportation}
\author{Ari Mizel}
\affiliation{Department of Physics, Pennsylvania State University, University Park, Pennsylvania 16802, U.S.A.}
\date{\today} 

\begin{abstract}
Ground-state quantum computers \cite{Mizel01,Mizel02} mimic quantum mechanical time evolution within the amplitudes of a time-independent quantum state.  We explore the principles that constrain this mimicking.  A no-cloning argument is found to impose strong restrictions.  It is shown, however, that there is flexibility that can be exploited using quantum teleportation methods to improve ground-state quantum computer design.
\end{abstract}

\pacs{03.67.Lx}

\maketitle

\section{I. Introduction}

The realization that quantum computers \cite{Nielsen00} can outperform classical computers on certain problems \cite{Shor94,Ekert96,Grover96,Grover97} has led to a surge of interest in the subject of quantum information theory.  On the one hand, abstract explorations have probed the characteristics and prospects of the theory \cite{Nielsen00,Shor94,Ekert96,Grover96,Grover97,Deutsch85,Jozsa91}.  On the other hand, experimental and theoretical research has pursued the realization of quantum information processing in the laboratory \cite{Cirac95,Monroe95,Turchette95,Gershenfeld97,Chuang98,Loss98,Burkhard99,Shnirman97,Makhlin99,Averin98,Nakamura99,Mooij99,Ioffe99,Vion02,Pashkin03,Kane98}.  There has been a spectrum of creative contributions in both directions.  Still, it remains unclear whether it will be feasible to develop a useful quantum computer, and it is also unclear what the potential of such a device ultimately is.  Given this situation, it is essential to continue exploring diverse approaches to this field, keeping in mind the compelling paradigms that have already emerged.

In previous articles, we have suggested a ``ground-state'' approach \cite{Mizel01,Mizel02} to quantum computing that departs from the conventional time-dependent picture.  In the usual picture of quantum computation, and indeed in general quantum mechanical time evolution, a system is characterized by a time-dependent state $\left| \psi (t_{i})\right>$ evolving as
\begin{equation}
\label{timeevolution}
\left| \psi (t_i)\right> = U_i \left| \psi (t_{i-1})\right>.
\end{equation}
In this equation, $t_j$ denotes a specific instant of time, with $j = 0,\dots,N$, and $U_j$ captures the evolution between $t_{i-1}$ and $t_i$.  The initial state of the system is $\left| \psi (t_{0})\right>$ and the final state, which presumably contains the results of the calculation, is $\left| \psi (t_{N})\right>$.
In ground-state quantum computation, the system is cooled into a stationary ground state $\left| \Psi \right>$ that has no time dependence.  Instead, the system is designed to have a large Hilbert space, so that all the quantum amplitudes in the entire sequence of states $\left\{\left|\psi(t_0) \right>,\left|\psi(t_1) \right>,\dots,\left|\psi(t_N) \right>\right\}$ are contained in $\left| \Psi \right>$.  In this way, the time evolution of $\left| \psi (t_{i})\right>$ is mimicked in a time-independent state.

\begin{figure}[tbp]
\includegraphics[height=5cm,angle=0]{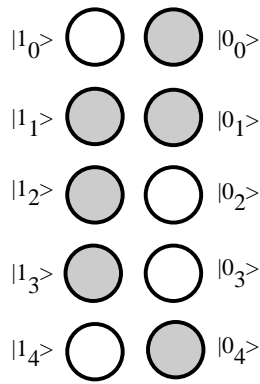} 
\caption{Quantum dot realization of ground state quantum computer with a single qubit.  An example is shown for $N=4$, where the electronic wavefunction (\ref{mimic}) purely for illustration is taken to be $\left[1\;\; 0\;\; \sqrt{1/2}\;\; \sqrt{1/2} \;\;0 \;\;1\;\; 0\;\; 1\;\; 1\;\; 0\right]^\dagger/\sqrt{5}$ and the shading indicates non-zero expectation value of the electronic charge density.}
\label{dots}
\end{figure}

For instance, in the case of a single qubit, there are two amplitudes in each $\left| \psi (t_i)\right>$, leading to a total of $2(N+1)$ amplitudes.  The ground-state quantum computer (GSQC) is therefore constructed with a $2(N+1)$ dimensional Hilbert space of states $\left\{\left| 0_0\right>,\left|1_0\right>,\left| 0_1\right>,\left|1_1\right>,\dots,\left| 0_N\right>,\left|1_N\right> \right\}$. The state $\left| \Psi \right>$ takes the following form when written as a column vector 
\begin{eqnarray}
\left[\!\! \begin{array}{c} \left< 0_0 \right. \left| \Psi \right> \\ \left< 1_0 \right. \left| \Psi \right> \\ \\ \left< 0_1 \right. \left| \Psi \right> \\ \left< 1_1 \right. \left| \Psi \right> \\ \vdots \\ \left< 0_N \right. \left| \Psi \right> \\ \left< 1_N \right. \left| \Psi \right> \end{array} \!\!\right] & = & \frac{1}{\sqrt{N+1}}\left[\!\! \begin{array}{c} \left< 0 \right. \left| \psi(t_0) \right> \\ \left< 1 \right. \left| \psi(t_0) \right> \\ \\ \left< 0 \right. \left| \psi(t_1) \right> \\ \left< 1 \right. \left| \psi(t_1) \right> \\ \vdots \\ \left< 0 \right. \left| \psi(t_N) \right> \\ \left< 1 \right. \left| \psi(t_N) \right> \end{array} \!\!\right] \nonumber \\
& = &\frac{1}{\sqrt{N+1}} \left[\!\! \begin{array}{r} \left[ \begin{array}{c} \left< 0 \right. \left| \psi(t_0) \right> \\ \left< 1 \right. \left| \psi(t_0) \right> \end{array} \right] \\ \\ U_1 \left[ \begin{array}{c} \left< 0 \right. \left| \psi(t_0) \right> \\ \left< 1 \right. \left| \psi(t_0) \right> \end{array} \right] \\ \hspace{0.2in} \vdots \hspace{0.4in} \\ U_N \dots U_1 \left[ \begin{array}{c} \left< 0 \right. \left| \psi(t_0) \right> \\ \left< 1 \right. \left| \psi(t_0) \right> \end{array} \right] \end{array}\!\! \right]
\label{mimic}
\end{eqnarray}
Equation (\ref{timeevolution}) has been invoked, and we see that the amplitudes contained in $\left| \Psi \right>$ depend upon the initial state $\left| \psi(t_0) \right>$ and the $2\times 2$ matrices $U_1,\dots,U_N$.  The particular physical realization of the state $\left| \Psi \right>$ is left unspecified in this formalism, just as the formalism of time-dependent quantum computation leaves the particular realization of the state $\left| \psi(t) \right>$ unspecified.  Experimental considerations would determine the best system for a GSQC apparatus.  For illustration purposes, it can be helpful to consider a single electron shared among $2(N+1)$ quantum dots, assuming one state per dot, as in Fig. \ref{dots}.

\begin{figure}[tbp]
\includegraphics[width=7cm,angle=0]{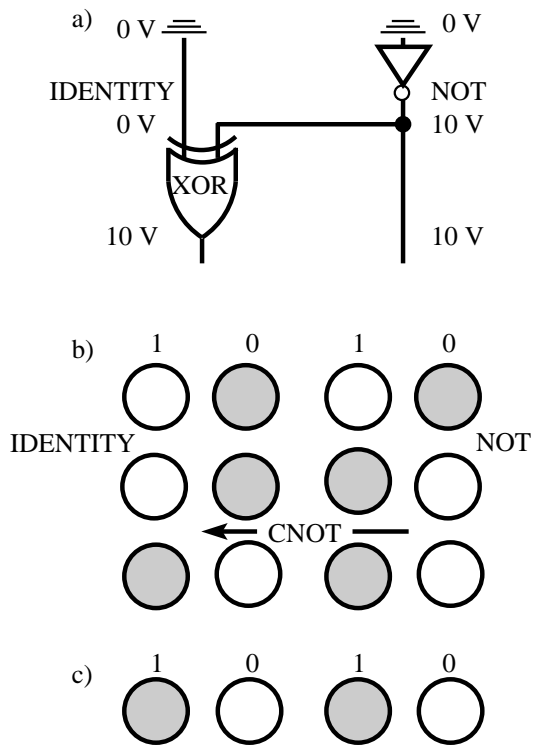} 
\caption{Comparison of (a) classical digital circuit, (b) GSQC implemented in an array of quantum dots, and (c) conventional time-dependent quantum computer realized with charge-based quantum dot qubits.  In (a), logical value 1 corresponds to $10$ V.  In (b), the expectation value of the ground-state charge density is indicated by shading of the quantum dots; it is analogous to the pattern of voltages in (a).  In (c), an electron shifts in time coherently between two dots in each qubit.  Only the final state is depicted.}
\label{comparetodigital}
\end{figure}

At first the ground-state scheme may appear unfamiliar, but it actually has a lot in common with the design of a classical digital computer.  In today's classical digital computers, during a given clock cycle, the electrical voltage establishes a time-independent, steady-state pattern in an array of gates. The input and output (and intermediate logical states) are simultaneously present as voltages at different spatial locations in the electric circuit.  An analogous situation prevails in a GSQC, in which a spatially extended quantum state plays the role of the electrical voltage pattern, achieving a time-independent state in an array of gates.  The comparison is illustrated in Figs. \ref{comparetodigital}a and \ref{comparetodigital}b.  The figures depict two bits which start an algorithm with logical value 0.  The left bit undergoes an IDENTITY and the right bit undergoes a NOT gate.  The two bits then undergo an XOR operation (CNOT in the quantum case, where the arrow points to the target bit), so that both finish with logical value 1.  In both the classical digital circuit and the GSQC, different steps in the logical flow correspond to different points in space.

In contrast, the conventional time-dependent quantum computer design actually functions quite differently than a classical digital computer (even while running a completely classical algorithm).  Instead of a steady-state pattern of voltage extended in space through an array of gates, one has time-dependent bits localized in space, subjected to time-dependent gates, as in Fig. (\ref{comparetodigital}c).  Classical computers are not made in this way -- there are significant design disadvantages involved in having the gates go to the bits rather than the bits go to the gates.

The compelling analogy between today's high performance digital circuits and GSQCs makes the prospect of constructing a GSQC in the laboratory seem more plausible.  A GSQC possesses additional favorable characteristics, as well, including an energy gap that defends against decoherence \cite{Mizel01}.  These attractive features are encouraging, but it is important to make a thorough and sober analysis of the principles that limit GSQCs.  A previous paper considered the challenges that arise with respect to computer scalability \cite{Mizel02}.  This paper takes a more comprehensive approach, investigating how much flexibility is possible when one seeks to mimic time evolution (\ref{timeevolution}) in a time-independent state $\left| \Psi \right>$.  Are we forced to use a state of form (\ref{mimic}), or are other possibilities available?  We show that the impossibility of cloning quantum information imposes fundamental restrictions upon the formalism, ruling out tensor product replacements of (\ref{mimic}).  On the other hand, the form (\ref{mimic}) is not completely without flexibility.  We show that it is possible to change (\ref{mimic}) with ``non-unitary'' gates.  Moreover, we show that non-unitary gates can be used in conjunction with quantum teleportation protocols to qualitatively improve GSQC design.

The paper is organized as follows.  Section II reviews the GSQC formalism and its scalability properties.  Section III, presents a spatial no-cloning argument that constrains the mimicking of time evolution.  This argument shows that it is not possible improve scalability by utilizing a tensor product version of (\ref{mimic}), which might seem attractive at first.  In section IV, we consider the flexibility that does exist in the formalism of ground-state quantum computation. Both many-particle qubits and non-unitary flexibility are explored.  Finally, in section V, we show that sidesteping the usual time-evolution with quantum teleportation, in conjunction with non-unitary gates, leads to a qualitative improvement in GSQC scalability.

\section{II. GSQC Hamiltonian and Scalability}

A single qubit system can be prepared in the state (\ref{mimic}) by setting up the following Hamiltonian on the $2(N+1)$ dimensional Hilbert space
\begin{equation}
\label{Hamiltonian}
H = \epsilon \left[ \begin{array}{ccccc} I + \Delta \sigma_z & -{U_1}^\dagger & & & \\ -U_1 & 2 I & - {U_2}^\dagger & & \\ & -U_2 & 2 I & - {U_3}^\dagger & \\ & & &\ddots & - {U_N}^\dagger \\ & & & -U_N & I \end{array} \right].
\end{equation}
Here, $I$ denotes the $2 \times 2$ identity matrix, $\sigma_z$ is a Pauli matrix, $\Delta$ is a small constant, and $\epsilon$ is some constant energy.  If $\Delta = 0$, the Hamiltonian (\ref{Hamiltonian}) is positive semi-definite and has two degenerate ground states of the form (\ref{mimic}).  (There are two states of the form (\ref{mimic}) since one can choose the input amplitudes $\left< 0 \right. \left| \psi(t_0) \right>$ and $\left< 1 \right. \left| \psi(t_0) \right>$ in two orthogonal ways.)  Fixing $\Delta$ at a small value introduces a small perturbing bias that selects out a unique ground state.  If $\Delta > 0$, the state (\ref{mimic}) with input $\left< 0 \right. \left| \psi(t_0) \right> = 0$ and $\left< 1 \right. \left| \psi(t_0) \right> = 1$ will be the Hamiltonian's unique ground state.  If $\Delta < 0$, the state (\ref{mimic}) with input $\left< 0 \right. \left| \psi(t_0) \right> = 1$ and $\left< 1 \right. \left| \psi(t_0) \right> = 0$ will be the Hamiltonian's unique ground state.  By setting up the Hamiltonian with a finite $\Delta$ and then cooling the system to a temperature that is low compared to the (free) energy of the excited states, one should be able to keep the system reliably in a unique ground state $\left| \Psi \right>$ with desired input.  

This Hamiltonian (\ref{Hamiltonian}) is written in second quantized notation in Ref. \cite{Mizel01}, as is its generalization for the case of multiple qubits.  In the single qubit case, the Hamiltonian (\ref{Hamiltonian}) can be rewritten 
\begin{equation}
H = \Delta C^{\dagger}_{0} \sigma_z C_{0} + \sum_{i=1}^{N} h^{i}(U_{i})
\end{equation}
where
\begin{equation}
\label{Hamsingle}
h^{i}(U_i) \! = \epsilon \!\!
 \left[
 C^{\dagger}_{i-1}C_{i-1} + C^{\dagger}_{i}C_{i} -
 \left(C^{\dagger}_{i} U_i C_{i-1} + {\rm h.c.}\right)\right]
\end{equation}
and $C^{\dagger}_{i} = \left[c^{\dagger}_{i,0}\;\; c^{\dagger}_{i,1}\right]$ is a column vector that groups together two creation operators $c^{\dagger}_{i,0}$ and $c^{\dagger}_{i,1}$ associated with states $\left| 0_i\right>$ and $\left| 1_i \right>$ respectively.  Although this is written in second quantized notation, there is only a single electron present in the single qubit case and its Hilbert space has dimension $2(N+1)$. 

 In the two qubit case, there are two electrons, each in a Hilbert space of dimension $2(N+1)$, so that the total Hilbert space has dimension $(2(N+1))^2$.  We add a qubit index $A$ or $B$ to all operators, and the Hamiltonian simply becomes
\begin{eqnarray}
H & = & \Delta_A C^{\dagger}_{A,0} \sigma_z C_{A,0} + \Delta_B C^{\dagger}_{B,0} \sigma_z C_{B,0} \nonumber \\
& & + \sum_{i=1}^{N} h^{A,i}(U_{A,i}) + h^{B,i}(U_{B,i})
\end{eqnarray}
assuming that each step $i$ in the algorithm stipulates independent single-qubit gates $U_{A,i}$ and $U_{B,i}$.  If the algorithm specifies as the $j$th operation a controlled-NOT of qubit $B$ by qubit $A$ instead of separate single-qubit gates, then we replace the terms $h^{A,j}(U_{A,j}) + h^{B,j}(U_{B,j})$ in $H$ with
\begin{eqnarray}
\label{CNOT}
h^{j}_{A,B}({\rm CNOT}) &=&
\epsilon  C_{A,j-1}^{\dagger}C_{A,j-1}C_{B,j}^{\dagger}C_{B,j} \\ \nonumber
&&+ h_{A}^{j}(I) C_{B,j-1}^{\dagger}C_{B,j-1}  \\ \nonumber
&&+ c_{A,j,0}^{\dagger}c_{A,j,0} h_{B}^{j}(I) \\ \nonumber
&&+  c_{A,j,1}^{\dagger}c_{A,j,1} h_{B}^{j}(N).
\end{eqnarray}

Given single-qubit gates (\ref{Hamsingle}) and the controlled-NOT gate (\ref{CNOT}), it follows that a GSQC can execute any quantum computation algorithm, on an arbitrary number of qubits \cite{Barenco95}.  In other words, ground-state quantum computation can essentially mimic arbitrary time evolution within the amplitudes of a state $\left| \Psi \right>$.

While this is satisfying from a formal standpoint, if one wishes to implement such a computer, a crucial concern is the scalability \cite{Mizel02}: the number of evolution steps that can be mimicked or, equivalently, the maximum length $N$ of the calculations executable on a GSQC.  At least three criteria determine scalability.  (i) The energy gap between the state of form (\ref{mimic}) with input value $0$ (i.e. $\left< 0 \right. \left| \psi(t_0) \right> = 1$ and $\left< 1 \right. \left| \psi(t_0) \right> = 0$) and the state of form (\ref{mimic}) with input value $1$ (i.e. $\left< 0 \right. \left| \psi(t_0) \right> = 0$ and $\left< 1 \right. \left| \psi(t_0) \right> = 1$) cannot decrease quickly with $N$.  This energy gap depends upon the fact that $\Delta \ne 0$ in $H$.  (ii) The energy gap between the ground state (\ref{mimic}) and excited states not of the form (\ref{mimic}) cannot decrease quickly with $N$.  It is essential that the gaps referred to in (i) and (ii) decrease slowly (or ideally stay constant) because we need to be able to cool the system to the ground state and keep it there reliably.  Assuming that the temperature of the computer can only be lowered to some given minimum value, we do not want thermal effects to overcome the gap and excite the computer out of its computationally meaningful ground state.  In addition, scalability requires that (iii) it be possible to extract the output of the calculation successfully from $\left| \Psi \right>$ with a probability not decreasing quickly with $N$.

Considering goals (i) and (ii), we determine the excitation spectrum of (\ref{Hamiltonian}).  By making the unitary transformation
\begin{equation}
\label{unitary}
U = \left[ \begin{array}{ccccc} I & & & & \\ & U_1 & & & \\ & & U_2 U_1 & & \\ & & &\ddots & \\ & & & & U_N \dots U_1 \end{array} \right] 
\end{equation}
we find that
\begin{equation}
U^\dagger HU = \epsilon \left[ \begin{array}{ccccc} I + \Delta \sigma_z & -I & & & \\ -I & 2 I & -I & & \\ & -I & 2 I & - I & \\ & & &\ddots & -I \\ & & & -I & I \end{array} \right].
\end{equation}
This is just the Hamiltonian of two uncoupled linear chains of length $N+1$ each, where the first site of one chain is perturbed by onsite energy $+\epsilon \Delta$ and the first site of the other chain by $-\epsilon \Delta$.  As far as (i) is concerned, we find that in first order perturbation theory the energies of the ground states of the two chains differ by $2\epsilon \Delta/(N+1)$ as a result of the perturbation.  To address (ii), the eigenspectrum of this Hamiltonian (with $\Delta$ set to zero) is calculated \cite{Mizel02}.  The gap to the first excited state of each chain is of order $\epsilon/(N+1)^2$, just as the eigenspectrum of a one dimensional box of length $L$ has energies scaling as $1/L^2$.  Thus, the smaller of the two gaps (i) and (ii) shrinks as $\epsilon/(N+1)^2$, which will impose a limit on the maximum length of a GSQC computation.

Naturally, it is worth considering whether (\ref{Hamiltonian}) could be replaced with another Hamiltonian with gaps that decrease slower with $N$ to improve scalability.  Certainly, there are many other positive-semidefinite Hamiltonians that have $\left| \Psi \right>$ as their ground state.  For instance, any power of the matrix (\ref{Hamiltonian}) will have this property.  In general, such Hamiltonians possess matrix elements involving products of the $U_i$, though (e.g. consider the form of $H^2$).  This is a great disadvantage since it would be necessary to compute these products classically in order to realize the Hamiltonian.  Such classical computations would be self-defeating -- in a sense, the very purpose of the quantum computation is to evaluate products of unitary matrices $U_i \dots U_1$ quantum mechanically.  Thus, (\ref{Hamiltonian}) seems to be especially appropriate for implementing a quantum algorithm with given input and $U_i$ but no additional information.

\section{III. Spatial No-Cloning}

Inspired in part by the scalability question, it is sensible to consider systematically how much flexibility is possible in the state $\left| \Psi \right>$.  The gap decreases with $N$ because the qubit wavefunction spreads out over a Hilbert space of increasing dimension $2(N+1)$.  Might it be possible to improve scalability by using many particles in small Hilbert spaces rather than a single particle in a large Hilbert space?  For instance, perhaps one could mimic time evolution of a qubit using a chain of $(N+1)$ spin-1/2 particles rather than the scheme of Fig. (\ref{dots}).

It seems reasonable to assume that in any time-mimicking framework, $\left| \Psi \right>$ will need to contain information about each of the time steps, $\left| \psi (t_0) \right>$, $\left| \psi (t_1) \right>$, $\dots$, $\left| \psi (t_N) \right>$.  In this case, the stationary state $\left| \Psi \right>$ must either contain a tensor product of the steps or a superposition of the steps; these are the only two ways to combine states in quantum mechanics.  A tensor product
\begin{equation}
\label{tensorproduct}
\left| \Psi \right> = \left| \psi (t_0) \right> \left| \psi (t_1) \right> \dots \left| \psi (t_N) \right>
\end{equation}
involves many bodies and potentially has desireable scalability properties.  However, we now argue that this form does not permit the enforcement of the desired connection (\ref{timeevolution}) between $\left| \psi (t_{i-1}) \right>$ and $\left| \psi (t_i) \right>$.  To see this, we make an argument along the lines of the no-cloning result of \cite{Wootters82}.  Consider a trivial computation that just clones its input as output, so that all of the $U_i$ in (\ref{timeevolution}) are identity operators.  Naturally, we will need to change the Hamiltonian that gives rise to $\left| \Psi \right>$ depending upon the value of the input $\left| \psi (t_0) \right>$.  However, it is reasonable to demand that the change be minor in some sense -- we do not want to have to embark on a ``pre-calculation'' to determine the Hamiltonian with the desired computationally meaningful ground state.  (In case of (\ref{Hamiltonian}) above, one just shifts the sign of $\Delta$ to change the input from 0 to 1.  Supplementing the calculation with a single qubit gate prior to the beginning of the algorithm even permits the input of an arbitrary superposition of 0 and 1.)  Clearly the minor change criterion is imprecise, and so the following must be regarded as just a plausibility argument.

Let us focus on the case of a single qubit.  If the computer performs this trivial cloning algorithm for input ``0,'' the state $\left| \Psi \right>$ is just \mbox{$\left| 0 \right> \left| 0 \right> \dots \left| 0 \right>$}.  For input ``1,'' the state $\left| \Psi \right>$ is just \mbox{$\left| 1 \right> \left| 1 \right> \dots \left| 1 \right>$}.  Presumably, these two states are nearly degenerate ground states of the Hamiltonian, and no minor change of the Hamiltonian can produce a major change in the states.  The minor change associated with selecting the input will just lead to a ground state \mbox{$\alpha \left| 0 \right> \left| 0 \right> \dots \left| 0 \right> + \beta \left| 1 \right> \left| 1 \right> \dots \left| 1 \right>$}.   In particular, if $\left| \psi (t_0) \right> =\frac{1}{\sqrt{2}}( \left| 0 \right> + \left| 1 \right>)$, then $\left| \Psi \right>$ will not take the desired form \mbox{$\frac{1}{\sqrt{2}}(\left| 0 \right> + \left| 1 \right>)\frac{1}{\sqrt{2}}(\left| 0 \right> + \left| 1 \right>)\dots \frac{1}{\sqrt{2}}(\left| 0 \right> + \left| 1 \right>)$}.  Instead, it will look something like \mbox{$\frac{1}{\sqrt{2}}\left| 0 \right> \left| 0 \right> \dots \left| 0 \right> + \frac{1}{\sqrt{2}} \left| 1 \right> \left| 1 \right> \dots \left| 1 \right>$}, a highly entangled state rather than a simple echoing of input to output.  In order to get the desired state, a major change in the Hamiltonian seems necessary.

We conclude that the tensor product form (\ref{tensorproduct}) is not suitable for a GSQC.  In a classical digital circuit, it is possible to use a voltmeter simultaneously to probe the value of the voltage at more than one point in the flow of logic, say both at the input and at the output of a given gate.  In a quantum computer, the initial state cannot coexist with the final state; unless the initial state is lost when the final state emerges from the gate, the two states end up entangled in an undesired fashion.  As above, if one imagines implementing an IDENTITY gate, the coexistence of the initial and final states would constitute cloning, clashing with the no-cloning result \cite{Wootters82}.

This is why it has become conventional to design a quantum computer that uses time-dependent localized bits.  These bits either experience an explicitly time-dependent Hamiltonian or traverse a time-independent Hamiltonian like mice in a maze (as in the ``flying qubit'' \cite{Turchette95} and ``cursor Hamiltonian'' \cite{Feynman85,Peres85} approaches).  With the passage of time, the initial state automatically evolves into the final state, so that the cloning problem is avoided.  If, instead of the time-dependent approach, one attempts to make a spatially extended quantum computer in analogy to the voltage pattern of a digital circuit, the cloning problem must be handled with more subtlety.  In (\ref{tensorproduct}), the states $\left| \psi (t_i) \right>$ are accessible to measurement for all values of $i$, and they become entangled with the final state $\left| \psi (t_N) \right>$ and frustrate quantum computation. The state (\ref{mimic}) sidesteps the cloning problem because the computer is placed into a {\em superposition} of initial and final states, so that both states are both present but cannot be simultaneously probed.  Design improvement must be pursued within this framework rather than using a tensor product (\ref{tensorproduct}).  

\section{IV. Flexibility in Mimicking Time Evolution}

\subsection{A. Many-particle Qubit}

A tensor product does not permit the mimicking of quantum time evolution, but it is still possible to design a GSQC qubit using a many-particle state.  Instead of realizing the Hamiltonian (\ref{Hamiltonian}) using a single particle in a $2(N+1)$ dimensional Hilbert space as in Fig. {\ref{dots}, one can set up a suitable $2(N+1)$ dimensional subspace of many-particle states.  For an illustrative example, consider a row of 3 quantum dots sharing a single electron, where we assume one state per dot.  We associate one of the dots with logical value 0, another dot with logical value 1, and the third dot with an ``idle'' condition.  The creation operators of the three states are $c^\dagger_0$, $c^\dagger_1$, and $d^\dagger$, respectively.  If $(N+1)$ of these arrangements are placed together, then Hilbert space has dimension $3^{(N+1)}$.  However, if the computationally meaningful states are those with $N$ electrons idle and only 1 ``non-idle,'' then the computationally meaningful Hilbert space has dimension $2(N+1)$.  These computationally meaningful states take the form $\left| 0 _i\right> = d^\dagger_{N} \dots d^\dagger_{i+1} c^\dagger_{0,i} d^\dagger _{i-1} \dots d^\dagger_0 \left| \mbox{vac}\right>$ or $\left| 1 _i\right> = d^\dagger_{N} \dots d^\dagger_{i+1} c^\dagger_{1,i} d^\dagger _{i-1} \dots d^\dagger_0 \left| \mbox{vac}\right>$.  One such state is depicted in Fig. \ref{idledots}.
\begin{figure}[tbp]
\includegraphics[width=8cm,angle=0]{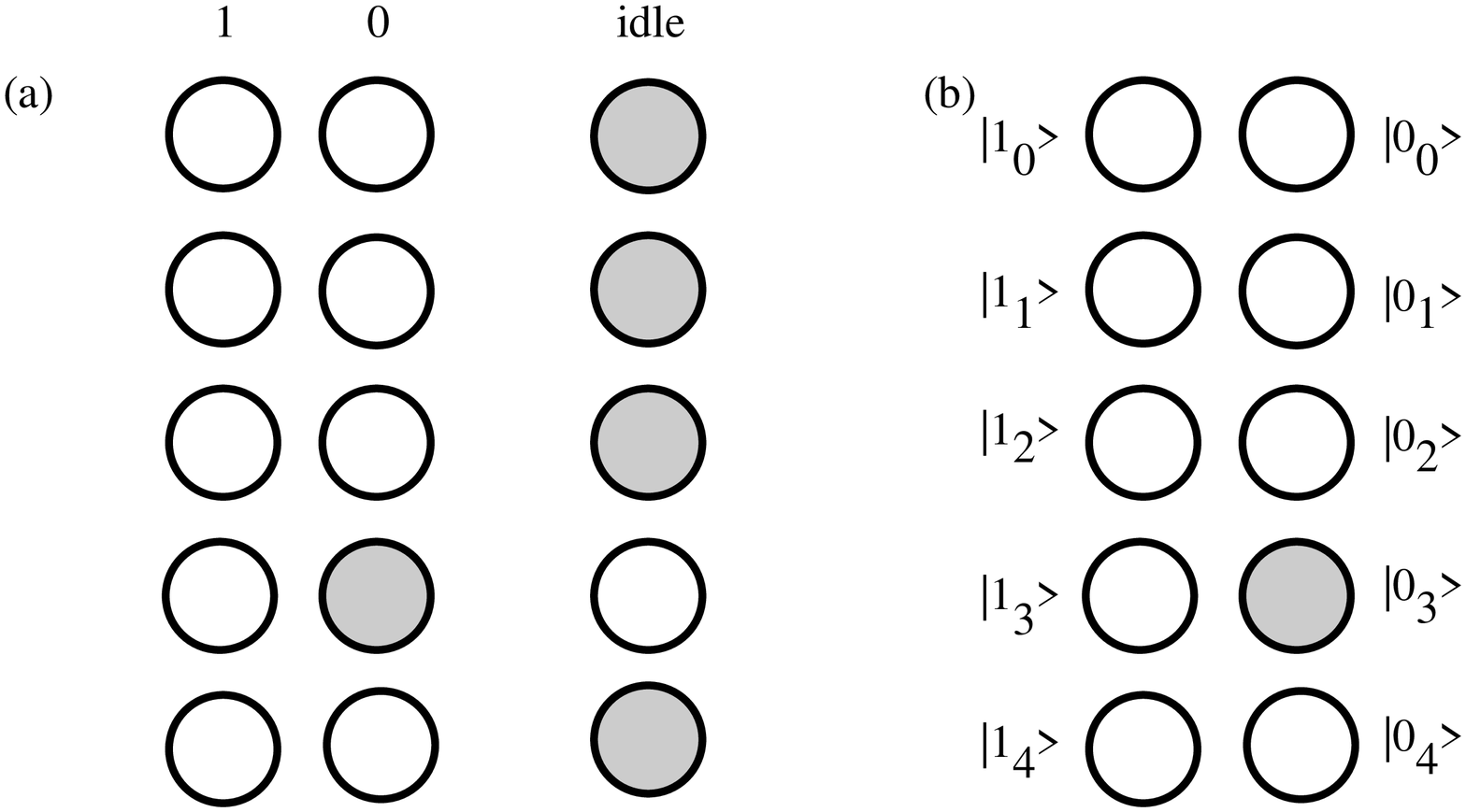} 
\caption{(a) A many-body realization of a single qubit.  Five ((N+1)=5 in this example) electrons are each confined to a row of $3$ dots.  The total Hilbert space has dimension $3^{N+1}=3^5$, but only $2(N+1) = 10$ states are computationally meaningful.  One such state, $\left| 0_3 \right>$, is portrayed here.  The state has $4$ electrons in their ``idle'' states, and electron $3$ in the logical 0 state.  (b) The same state in the single electron implementation of Fig. \ref{dots}.}
\label{idledots}
\end{figure}
With this many-body realization of $\left| 0 _i\right>$ and $\left| 1_i\right>$, state (\ref{mimic}) is the ground state of the Hamiltonian $\tilde{H} = \Delta \; C^{\dagger}_{0} \sigma_z C_{0} + \sum _{i=1}^N \tilde{h}^i(U_i)$ where
\begin{equation}
\label{singlequbit}
\tilde{h}^i(U) = \epsilon \left[ C^\dagger_{i} C_{i} + C^\dagger_{i-1} C_{i-1} - (d^\dagger _{i-1} C^\dagger_{i} U_i C_{i-1} d_{i} + H.c.) \right]
\end{equation}
and the subscript $i$ distinguishes among the $N+1$ rows, each carrying a single electron. We have grouped together the non-idle states into a column vector $C_i^\dagger \equiv \left[ c^\dagger _{i,0} \; c^\dagger _{i,1} \right]$ as in (\ref{Hamsingle}).   In this many-body implementation of a qubit, even single-qubit gates require two-body interactions because gate $i$ must scatter the electron in row $i-1$ into its idle state and scatter the electron in row $i$ out of its idle state into a logical state 0 or 1.

The extension to the case of two qubits $A$ and $B$ requires that we attach a qubit label to each operator and write $\tilde{H} = \Delta_A C^{\dagger}_{A,0} \sigma_z C_{A,0} + \Delta_B C^{\dagger}_{B,0} \sigma_z C_{B,0} + \sum_{i=1}^{N} \tilde{h}^{A,i}(U_{A,i}) + \tilde{h}^{B,i}(U_{B,i})$.  Suppose that an algorithm specifies as operation $j$ a controlled-NOT of qubit $B$ by qubit $A$ rather than independent single qubit gates.  Then the terms $\tilde{h}^{A,j}(U_{A,j}) + \tilde{h}^{B,j}(U_{B,j})$ are removed from the Hamiltonian.  Single qubit gates (\ref{singlequbit}) require two-body interactions in this many-body implementation, and proceeding in a direct manner, we might be tempted to devise a controlled-NOT gate that involves unphysical four body interactions.  To avoid this, the controlled-NOT is implemented in the same way as in the previous GSQC implementation.  The control qubit's row $j-1$ electron is allowed to inhabit 2 rows of logical quantum dots instead of just 1 row.  The same is done for the target qubit (see Fig. \ref{idledotsCNOT}).  Thus, each row $j-1$ electron occupies $5$ quantum dots instead of $3$, including the two states grouped into $C^\dagger _{A,j-1} \equiv \left[ c^\dagger _{A,j-1,0} \; c^\dagger _{A,j-1,1} \right] $, the idle state $d^\dagger _{A,j-1}$, and also the two non-idle states grouped into in $C^\dagger_{A,j}\equiv \left[ c^\dagger _{A,j,0} \; c^\dagger _{A,j,1} \right]$.  
\begin{figure}[tbp]
\includegraphics[width=9cm,angle=0]{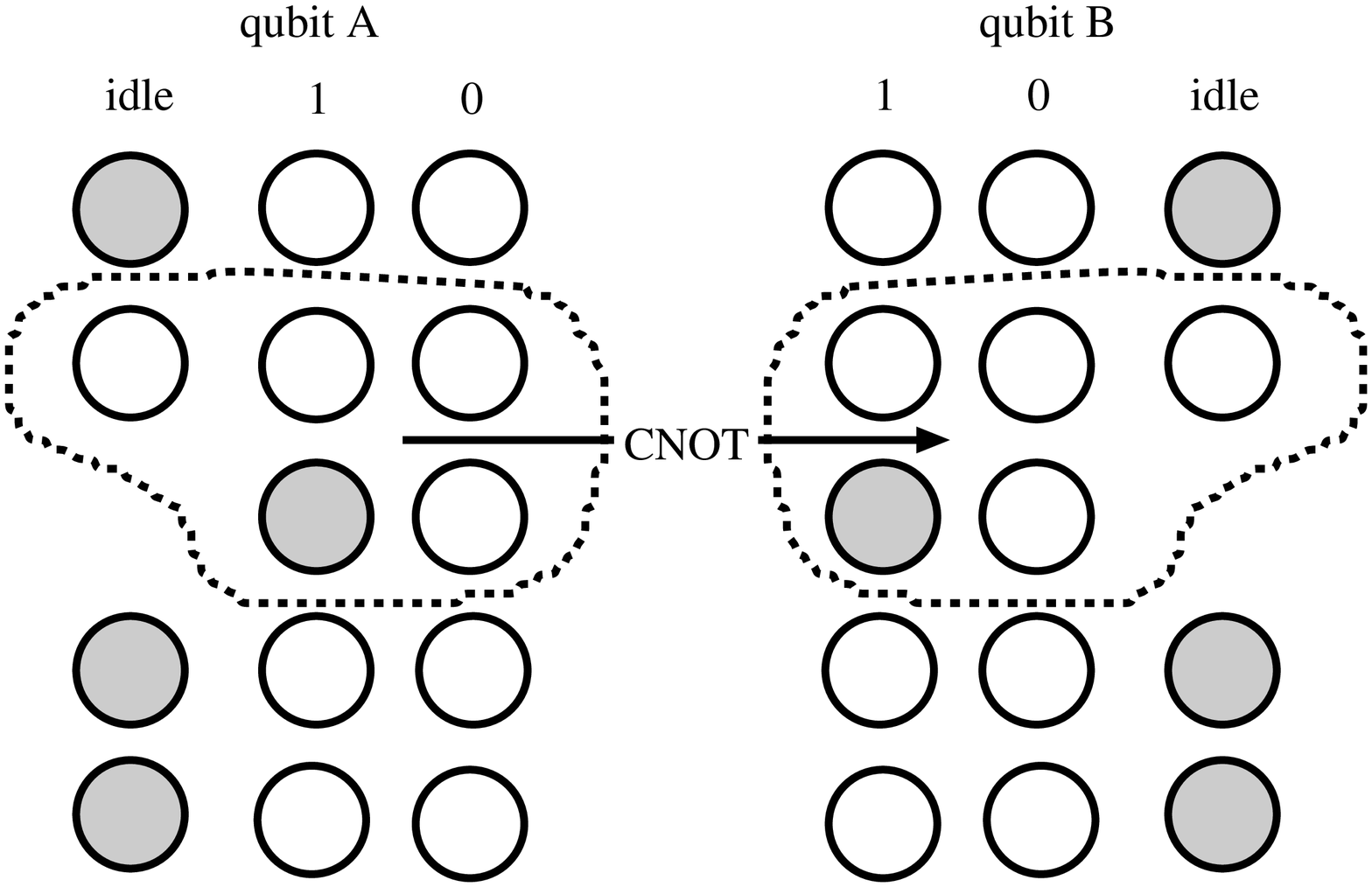} 
\caption{Controlled-NOT in many-body-qubit realization of a GSQC.  Each row of $3$ quantum dots in qubit $A$ contains a single electron, as does each row of $3$ quantum dots in qubit $B$.  Within each dotted line, however, there are $5$ dots instead of $3$ that contain a single electron.  The computer executes a controlled-NOT between between row $j=1$ and row $j=2$ using the Hamiltonian (\ref{CNOT}) just as in the previous implementation.}
\label{idledotsCNOT}
\end{figure}
The term $h^j_{A,B}(CNOT)$ of exactly form (\ref{CNOT}) is added to the Hamiltonian.  After the controlled-NOT, the algorithm resumes with step $j+1$ for the control qubit by adding to the Hamiltonian a slight modification of (\ref{singlequbit}) 
\begin{eqnarray}
\lefteqn{ \tilde{\tilde{h}}^{A,j+1}(U_{A,j+1})  =  \epsilon \left[ C^\dagger_{A,j+1} C_{A,j+1} + C^\dagger_{A,j} C_{A,j} \right.} \nonumber \\
& & \left. - (d^\dagger _{A,j-1} C^\dagger_{A,j+1} U_{A,j+1} C_{A,j} d_{A,j+1} + H.c.) \right]
\end{eqnarray}
because there is no ``idle'' dot in row $j$.  The target qubit resumes by adding a similar term, with the label $A$ replaced by $B$.  Subsequent terms have the form (\ref{singlequbit}).

With single-qubit gates and the controlled-NOT gate in hand, we have a many-body implementation that constitutes a complete alternative to the single-electron-qubit realization of Fig. \ref{dots}.  This flexibility of implementation is noteworthy.  However, the many-body character of the state is not found to lead to any advantages. In the many-particle ground-state, electrons mostly just occupy their ``idle'' state (i.e. an arbitrary electron $i$ in an arbitrary qubit $M$ has $\left< \Psi \right| d^\dagger_{M,i}d_{M,i}\left| \Psi \right> \gg \left< \Psi \right| C^\dagger_{M,i}C_{M,i}\left| \Psi \right>$).   The Hamiltonian in the computationally meaningful Hilbert space of this system has the form (\ref{Hamiltonian}), so the spectrum is no better than the original spectrum as far as the scalability criteria (i) - (iii) of section II are concerned.  Other sources of flexibility in the formalism must be sought.

\subsection{B. ``Non-unitary'' Flexibility}

The form (\ref{mimic}) of $\left| \Psi \right>$ treats every time step equivalently, which is a characteristic of genuine time evolution, but is not essential when we are mimicking time evolution.  For instance, we may be more concerned with the output of a calculation, $U_N \dots U_1 \left[ \begin{array}{c} \left< 0 \right. \left| \psi(t_0) \right> \\ \left< 1 \right. \left| \psi(t_0) \right> \end{array} \right]$,  than with its intermediate steps.  We are led to consider the possibility that $\left| \Psi \right>$ takes the form
\begin{equation}
\label{mimiclambda}
\left[\!\! \begin{array}{c} \left< 0_0 \right. \left| \Psi \right> \\ \left< 1_0 \right. \left| \Psi \right> \\ \\ \left< 0_1 \right. \left| \Psi \right> \\ \left< 1_1 \right. \left| \Psi \right> \\ \vdots \\ \left< 0_N \right. \left| \Psi \right> \\ \left< 1_N \right. \left| \Psi \right> \end{array} \!\!\right] \!\! = \!\! \frac{1}{\sqrt{N+1}} \left[\!\! \begin{array}{r} \lambda_0 \left[ \begin{array}{c} \left< 0 \right. \left| \psi(t_0) \right> \\ \left< 1 \right. \left| \psi(t_0) \right> \end{array} \right] \\ \\ \lambda_1 U_1 \left[ \begin{array}{c} \left< 0 \right. \left| \psi(t_0) \right> \\ \left< 1 \right. \left| \psi(t_0) \right> \end{array} \right] \\ \hspace{0.2in} \vdots \hspace{0.4in} \\ \lambda_N U_N \dots U_1 \left[ \begin{array}{c} \left< 0 \right. \left| \psi(t_0) \right> \\ \left< 1 \right. \left| \psi(t_0) \right> \end{array} \right] \end{array} \!\!\right]
\end{equation}
where $\sum_{i=0}^N |\lambda_i| ^2 = N+1$.  This constitutes a ``non-unitary'' form of evolution, in which the overall probability is not conserved from step to step.  This $\left| \Psi \right>$ is the ground state of the Hamiltonian
\begin{equation}
\label{Hamiltonianpot}
H \! = \! \epsilon \! \left[\! \begin{array}{ccccc} v_0 (I+ \Delta \sigma_z) & -t_1^*U_1^\dagger & & & \\ -t_1 U_1 & 2 v_1 I & - t_2^* U_2^\dagger & & \\ & -t_2 U_2 & 2 v_2 I & - t_3^* U_3^\dagger & \\ & & &\ddots & - t_N^* U_N^\dagger \\ & & & -t_N U_N & v_N I \end{array} \!\! \right] 
\end{equation}
where $t_i$ and $v_i$ are numbers satisfying \mbox{$-t_{i} \lambda_{i-1} + 2 v_i \lambda _i - t_{i+1}^* \lambda _{i+1} = 0$} for $1\le i < N$, $v_0\lambda_0 - t_1 ^* \lambda_1 =0$, and $-t_N \lambda_{N-1} + v_N \lambda_N = 0$.  (We assume that $\epsilon$ has been chosen so that $|v_i| \le 1$ and $|t_i| \le 1$).  Using the unitary transformation (\ref{unitary}) we find
\begin{equation}
U^\dagger HU = \epsilon \left[ \begin{array}{ccccc} v_0 (I+ \Delta \sigma_z) & -t_1^* I & & & \\ -t_1 I & 2 v_1 I & -t_2^* I & & \\ & -t_2 I & 2 v_2 I & - t_3^*I & \\ & & &\ddots & -t_N^*I \\ & & & -t_N I & v_N I \end{array} \right].
\end{equation}
Thus, the spectrum is that of two uncoupled linear chains of length $N$,  as in Fig. \ref{chain}, where for each chain the onsite potential at position $i$ is $2\epsilon v_i$ and the tunneling matrix element between sites $i-1$ and $i$ is $\epsilon t_i$.  At the ends of each chain, the onsite potential is $\epsilon v_i$ rather than $2\epsilon v_i$, and a small perturbation $\pm \epsilon v_0 \Delta$ breaks the equivalence of the chains.
\begin{figure}[tbp]
\includegraphics[height=6cm,angle=0]{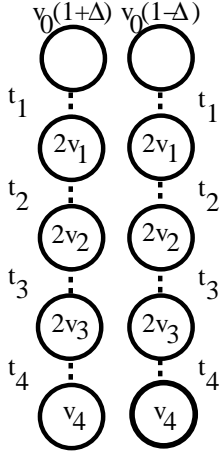} 
\caption{The spectrum of the Hamiltonian (\ref{Hamiltonianpot}) is that of two chains with onsite potentials $2 \epsilon v_i$ and tunneling matrix elements $\epsilon t_i$.  At the ends of each chain, the onsite potential is $\epsilon v_i$ rather than $2\epsilon v_i$, and a small perturbation $\pm \epsilon v_0 \Delta$ breaks the equivalence of the chains, so that the system has a single non-degenerate eigenstate.  The constant energy factor $\epsilon$ is omitted from the figure.}
\label{chain}
\end{figure}
Flexibility in the $t_i$ and $v_i$ can be used to tailor the $\lambda _i$.  

To satisfy scalability criterion (ii) above, one can set up a
pronounced minimum in the onsite potential $v_i$.  This can produce a
ground state in each chain with an energy gap that is finite in the $N \rightarrow
\infty$ limit. However, the ground state will be localized, with
values of $\lambda_i$ that decay rapidly away from the potential
minimum, as shown in Fig. \ref{threegoals}a.  To satisfy criterion (i), one can locate the minimum in the
potential $v_i$ near $i = 0$ to tailor $\lambda_0$ to be of order $N$, as in Fig. \ref{threegoals}b.
First order perturbation theory implies that the state of form
(\ref{mimiclambda}) with input value 0 will differ in energy by $2
\epsilon \Delta |\lambda _0|^2/(N+1)$ from the state of form
(\ref{mimiclambda}) with input value 1; if $\lambda_0$ is of order $N$
this energy difference will not decrease with $N$.  Alternatively, to
satisfy criterion (iii), one can locate the minimum in the potential $v_i$
near $i = N$ to tailor $\lambda_N$ to be of order $N$, as in Fig. \ref{threegoals}c. In the case
of classical output, the results of the calculation can be
measured with unit probability using sensor electrons \cite{Mizel02};
the gap of sensor electrons scales as $\epsilon |\lambda
_N|^2/(N+1)$.  For $\lambda_N$ of order $N$, this gap will not decrease with $N$.

\begin{figure}[tbp]
\includegraphics[height=6cm,angle=0]{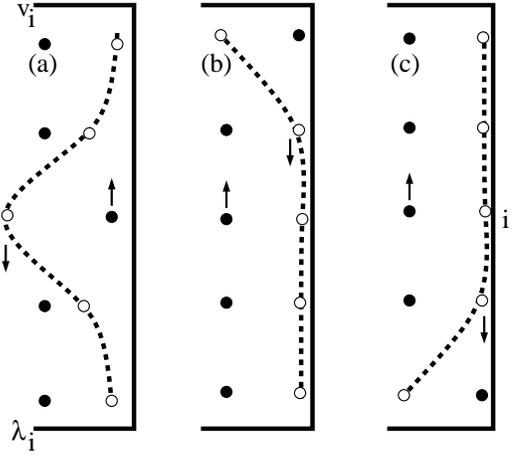} 
\caption{Three choices of $v_i$ and the corresponding values of $\lambda_i$.  In each case, the filled circles give values of $v_i$ and the empty circles give values of $\lambda_i$.  The dashed lines are guides to the eye of the form of the $\lambda_i$. (a) A minimum in the $v_i$ leads to an energy gap in accordance with scalability criterion (ii) but the $\lambda_i$ are substantial only in a localized region of $i$.  (b) If the minimum is near $i=0$, then $|\lambda_0|$ large but $|\lambda_N|$ is small, conflicting with (iii).  (c) If the minimum is near $i=N$, then $|\lambda_N|$ large but $|\lambda_0|$ is small, conflicting with (i).}  
\label{threegoals}
\end{figure}

Unfortunately, no single choice of $v_i$ and $t_i$ seems capable of
simultaneously satisfying all three criteria (i)-(iii).  The solution to criterion
(ii) involves a localized ground state that cannot have large values
for both $|\lambda _0|$ and $|\lambda_N|$, which are on opposite sides
of the chain.  Either $|\lambda_0|$ or $|\lambda_N|$ will be small,
conflicting with criterion (i) or (iii) respectively.

Of course, one can put two minima in $v_i$, so that the ground state
will be like the symmetric solution to a double well potential,
possessing large $|\lambda_0|$ and large $|\lambda_N|$.  However, in
this case the first excited state will simply be like the
antisymmetric solution of the double well potential, and there will be
a very small energy gap in violation of criterion (ii).  Attempts to
reduce $t_i$ (recall that $t_i$ cannot be increased because $|t_i| \le
1$ by definition) seem simply to reduce energy penalty associated with
putting nodes in the wavefunction, exacerbating the conflict with
criterion (ii).

\section{V. Gate Application by Quantum Teleportation}

The non-unitary character of (\ref{mimiclambda}) seems to be the primary source of flexibility in $\left| \Psi \right>$.  While directly tuning the $\lambda_i$ does not seem to improve GSQC scalability, we now show that a different approach is possible.  Non-unitary gates can be utilized to improve GSQC scalability when combined with the protocol of quantum teleportation \cite{Bennett93}.

Quantum teleportation can be used to apply gates \cite{Nielsen97,Gottesman99,Vidal02} by exploiting the following equality
\begin{eqnarray}
\label{teleportidentity}
\lefteqn{U_1 \left| 0 \right> \frac{\left|0\right> U_2 \left|0 \right>+\left|1\right> U_2 \left|1 \right>}{\sqrt{2}}} \nonumber \\
 & = & \frac{1}{2} \left[\left| \Phi _0 \right> U_2 \sigma_0 U_1 \left| 0 \right> + \left| \Phi _1 \right> U_2 \sigma_1 U_1 \left| 0 \right> \right. \nonumber \\
& & \left. + \left| \Phi _2 \right> U_2 \sigma_2 U_1 \left| 0 \right> + \left| \Phi _3 \right> U_2 \sigma_3 U_1 \left| 0 \right> \right] \nonumber \\
& = & \frac{1}{2} \sum_{i} \left| \Phi _{i} \right> U_2 \sigma_{i} U_1 \left| 0 \right>
\end{eqnarray}
where $\left| \Phi _i \right> = \frac{1}{\sqrt{2}} ( \left| 0\right > \sigma_i \left| 0 \right> + \left| 1 \right> \sigma_i \left| 1 \right>)$ and $\sigma_i$ is a Pauli matrix (with $\sigma_0 = I$).  This equality is most simply demonstrated by writing out $U_1 \left| 0 \right>$ explicitly as $a \left| 0 \right> + b\left| 1 \right>$.  In the case that $U_2$ is the identity operation, this equality is used to quantum teleport \cite{Bennett93} the state $U_1 \left| 0 \right>$ into the second qubit of an entangled EPR pair.

To see why this is useful for our purposes, we generalize (\ref{teleportidentity}) as follows
\begin{eqnarray}
\lefteqn{U_1 \left| 0 \right> \frac{\left|0\right> U_2 \left|0 \right>+\left|1\right> U_2 \left|1 \right>}{\sqrt{2}} \frac{\left|0\right> U_3 \left|0 \right>+\left|1\right> U_3 \left|1 \right>}{\sqrt{2}}}  \nonumber \\
& & \hspace{1.0in} \;\;\;\;\;\;\;\;\;\;\;\; \dots \frac{\left|0\right> U_N \left|0 \right>+\left|1\right> U_N \left|1 \right>}{\sqrt{2}} \nonumber \\
& = &  \frac{1}{2} \sum_{i} \left| \Phi _{i} \right> \left( U_2 \sigma_{i} U_1 \left| 0 \right>\frac{\left|0\right> U_3 \left|0 \right>+\left|1\right> U_3 \left|1 \right>}{\sqrt{2}} \right. \nonumber \\
& & \left. \hspace{1.0in} \;\;\;\;\;\;\;\;\;\;\;\; \dots \frac{\left|0\right> U_N \left|0 \right>+\left|1\right> U_N \left|1 \right>}{\sqrt{2}} \right) \nonumber \\
& = & \!\!\!\! \frac{1}{2^{N-1}} \!\!\!\!\!\!\!\! \sum _{i_1,\dots,i_{N-1}} \!\!\!\!\!\! \left| \Phi _{i_1} \right> \! \dots \! \left| \Phi _{i_{N-1}} \right> U_N \sigma_{i_{N-1}}\dots U_2 \sigma_{i_1} U_1 \!\! \left| 0 \right> \label{teleportstate}
\end{eqnarray}
This equation shows that we do not have to apply the unitary operators $U_i$ in series in order to produce the result $U_N \dots  U_1 \left| 0 \right>$.  Instead, we can apply the gates in parallel.

An explicit 7 step procedure to produce $U_N \dots  U_1 \left| 0 \right>$ is as follows.  (a) Initialize $2N-1$ qubits in logical 0, with overall state \mbox{$\left| 0 \right> \dots \left| 0 \right>$}, (b) apply a Walsh-Hadamard gate $W = \frac{1}{\sqrt{2}}\left[\begin{array}{lr} 1 & 1\\1 & -1 \end{array} \right]$ to every other qubit in parallel to obtain the state \mbox{$\left| 0 \right>\frac{\left| 0 \right> + \left| 1 \right>}{\sqrt{2}}\left| 0 \right>\frac{\left| 0 \right> + \left| 1 \right>}{\sqrt{2}}\dots \left| 0 \right>$}, (c) apply a controlled-NOT gate to every other pair of qubits to produce $N$ EPR pair states and one logical 0 in the state \mbox{$\left| 0 \right> \frac{\left|0\right>\left|0 \right>+\left|1\right>\left|1 \right>}{\sqrt{2}} \frac{\left|0\right>\left|0 \right>+\left|1\right>\left|1 \right>}{\sqrt{2}}\dots\frac{\left|0\right>\left|0 \right>+\left|1\right>\left|1 \right>}{\sqrt{2}}$}, (d) apply $U_i$ in parallel to every other qubit, yielding the state (\ref{teleportstate}).  Then, by measuring the $2N$ initial qubits, there is some probability that every pair will be found in the state $\left| \Phi _{0} \right>$ and the remaining qubit will be in the desired state.  The measurement can be accomplished by (e) executing a CNOT gate between adjacent pairs of qubits and then (f) executing a Walsh-Hadamard gate so that $\left| \Phi _{0} \right>$ becomes $\left|0 \right> \left| 0 \right>$, $\left| \Phi _{1} \right>$ becomes $\left|0 \right> \left| 1\right>$,$\left| \Phi _{2} \right>$ becomes $-i \left|1 \right> \left| 1\right>$, and $\left| \Phi _{3} \right>$ becomes $\left|1 \right> \left| 0\right>$.  If (g) the initial $2N$ qubits are all measured to be in the state $\left| 0 \right>$, then the final qubit will be in the desired state $U_N \dots  U_1 \left| 0 \right>$.  If we are working in the context of time-dependent quantum computation and only have unitary evolution at our disposal, then the probability of obtaining the correct result is $1/2^{2N-2}$, the square of the amplitude of the desired term in equation (\ref{teleportstate}).  However, if we are mimicking time evolution, it is possible to make the evolution non-unitary and to increase the probability of obtaining the correct result.  Since there are only 7 time steps (a) - (g) to be mimicked regardless of $N$, it turns out to be possible to improve computer scalability.

A GSQC design that implements these 7 steps for the case $N = 3$ is portrayed in Fig. \ref{teleportfig}.   The extension to larger $N$ is straightforward.
\begin{figure}[tbp]
\includegraphics[width=9cm,angle=0]{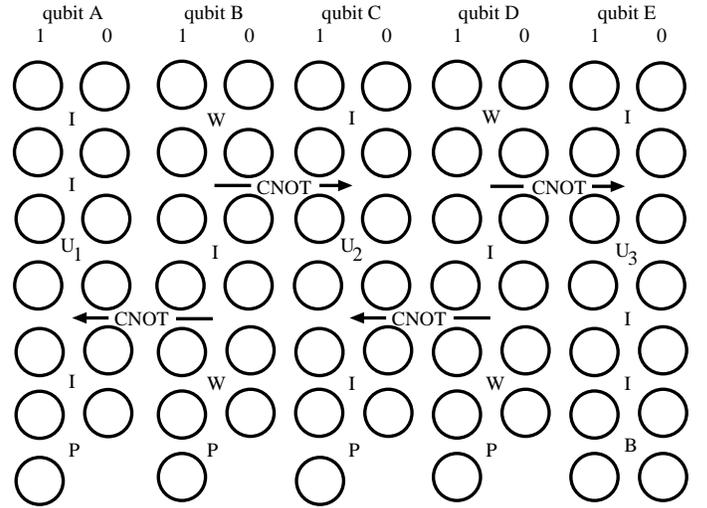} 
\caption{Ground-state computer that applies gates by quantum teleportation to produce $U_N\dots U_1\left|0\right>$.  Example is shown for $N=3$ operations, where the protocol entails 5 qubits.  The 7 rows of the computer execute steps (a)-(g) in sequence.  Rows 1-3 yield two EPR pairs, assuming that all qubits are input with logical 0.  Row 4 applies the unitary gates $U_i$.  Rows 5-7 measure 4 of the 5 qubits, and the desired state resides with some probability in the final row of qubit E.  To increase $N$, more qubits are added and the same pattern of gates is employed.}
\label{teleportfig}
\end{figure}
For the case $N=3$, the appropriate Hamiltonian has the form
\begin{eqnarray}
& & H = \Delta_A C^{\dagger}_{A,0} \sigma_z C_{A,0} + \dots + \Delta_E C^{\dagger}_{E,0} \sigma_z C_{E,0} \nonumber\\
& & + h^{A,1}(I) + h^{B,1}(W)+ h^{C,1}(I) + h^{D,1}(W)+ h^{E,1}(I) \nonumber \\
& & + h^{A,2}(I) + h^2_{B,C}(CNOT) + h^2_{D,E}(CNOT) \nonumber \\
& & + h^{A,3}(U_1) + h^{B,3}(I)+ h^{C,3}(U_2) + h^{D,3}(I)+ h^{E,3}(U_3) \nonumber\\
& & + h^4_{B,A}(CNOT) + h^4_{D,C}(CNOT) + h^{E,4}(I)  \nonumber \\
& & + h^{A,5}(I) + h^{B,5}(W)+ h^{C,5}(I) + h^{D,5}(W) + h^{E,5}(I) \nonumber \\
& & + h^{A,6}(P) + h^{B,6}(P) + h^{C,6}(P) + h^{D,6}(P) + h^{E,6}(B). \nonumber
\end{eqnarray}
Each line of this equation implements one of the steps (a) - (g) of the procedure.  The single qubit gate (\ref{Hamsingle}) and the controlled-NOT gate (\ref{CNOT}) are employed repeatedly.  For $\Delta_A,\dots,\Delta_E < 0$, the first line ensures that the ground state will have all qubits begin with logical 0, which is step (a).  The second line executes step (b), applying the requisite Walsh-Hadamard operations.  The third line effects the controlled-NOT operations that produce the EPR pairs, in accordance with step (c).  The fourth line carries out step (d), acting with the $U_i$ gates.  The fifth and sixth lines perform steps (e) and (f), taking the states $\left| \Phi_i\right>$ into product states in preparation for measurement.  At the final line, a new single-qubit Hamiltonian $h^{j}(P)$ causes non-unitary evolution.  It projects out the part of $\left| \Psi \right>$ that has $\left| 0 \right>$ values for qubits $A$,$B$,$C$, and $D$, and increases the amplitude of this part to enhance the chance of success in step (g).  This projection Hamiltonian takes the form
\begin{equation}
h^{i}(P) = \epsilon
 \left[c^{\dagger}_{i-1,0}c_{i-1,0} + \frac{1}{\lambda^2} c^{\dagger}_{i,0}c_{i,0} - \frac{1}{\lambda}
 \left(c^{\dagger}_{i,0}c_{i-1,0} + {\rm h.c.}\right)\right].
\end{equation}
for some constant $\lambda > 1$.  In addition, in the last line there is a boost Hamiltonian $h^{E,6}(B)$ that boosts the amplitude for the electron of qubit $E$ to be on the final line, but without a projection
\begin{equation}
h^{i}(B) \! = \epsilon \!\! \left[ C^{\dagger}_{i-1}C_{i-1} + \frac{1}{\lambda^2}C^{\dagger}_{i}C_{i} -\frac{1}{\lambda} \left(C^{\dagger}_{i} C_{i-1} + {\rm h.c.}\right)\right]. \end{equation}

By choosing this Hamiltonian in accordance with the 7 step procedure, we ensure that the ground state includes a desired contribution of the form $c^\dagger_{A,6,0}c^\dagger_{B,6,0}c^\dagger_{C,6,0}c^\dagger_{D,6,0}\left(C^\dagger_{E,6}U_3U_2U_1\left[\begin{array}{c}1\\0\end{array}\right]\right)\left| {\rm vac} \right>$, in which all 5 electrons are found in the final row of the computer.  There are many other terms in which at least one electron is not in its final row.  The probability of extracting the desired term by measuring all electrons in the final row is greater than $\left(\frac{\lambda^2/2}{6+\lambda^2/2}\right)^{2N-2}\frac{\lambda^2}{6+\lambda^2}=\left(\frac{\lambda^2/2}{6+\lambda^2/2}\right)^4\frac{\lambda^2}{6+\lambda^2}$ \cite{greaterthan}.  We compute this by nothing that the first $(2N-2) = 4$ electrons have probability greater than $\frac{\lambda^2/2}{6+\lambda^2/2}$ of making it through the projection and residing on the last row, while the final electron has probability greater than $\frac{\lambda^2}{6+\lambda^2}$ of residing on the two dots of the last row.  If we choose $\lambda \sim N$, then our probability of accessing the desired state will not decrease with $N$.

In this case of $\lambda \sim N$, we can estimate using the analysis of \cite{Mizel02} that the resulting gap of this computer should scale roughly as $\epsilon/(6(6+\lambda^2)) \sim \epsilon /N$.  This represents an important improvement over the our original form (\ref{mimic}) of reaching the state $U_N\dots U_1 \left| 0 \right>$ whose gap scaled as $\epsilon /(N+1)^2$.

This quantum teleportation means of applying GSQC gates can be extended to a computer with many qubits.  It is straightforward to include controlled-NOT gates, given the discussion here and the treatment in Ref. \cite{Gottesman99}.  Our gap estimate of $\sim \epsilon/N$ remains true in the multiple qubit case.  

\section{VI. Conclusion}

We have studied the mimicking of quantum mechanical time evolution (\ref{timeevolution}) within the amplitudes of a time-independent state $\left| \Psi \right>$.  A no-cloning type of principle imposes strong constraints upon the form of $\left| \Psi \right>$.  Nevertheless, important flexibility remains, especially in the form of non-unitary evolution.  We have demonstrated how this flexibility can be exploited, together with quantum teleportation methods, to improve GSQC design.  Consulting the three scalability criteria mentioned above, we note that the resulting GSQC has a gap (i,ii) that decreases as $\epsilon /N$ when (iii) the measurement probability is non-decreasing with $N$. The result is a much more scalable GSQC.

It is possible that a gap scaling of $\epsilon/N$ may constitute a fundamental maximum for any GSQC.  Suppose that the inverse gap were to set the time scale for settling into the ground state and thereby obtaining the answer to a calculation. If the inverse gap of some GSQC were less than $O(N)$, then the answer to an $N$ step calculation could be available in a time less than $O(N)$, which would be surprising.  To avoid this, it might be necessary that the gap of every GSQC be no less than $O(1/N)$.  Of course, this argument is quite heurisitic.

Irrespective of the size of the gap, it may be fruitful in future work to try to introduce a clock cycle into GSQC design in analogy to the clocking of classical digital computers.  For instance, it might be possible to shift the onsite potentials in (\ref{Hamiltonianpot}) adiabatically, so that the $v_i$ would exhibit a minimum at $i=0$ at the beginning of a calculation and the minimum would move slowly down the array until reaching $i=N$ at the end of the calculation.  This would sweep a localized ground state through the quantum dot array \cite{Mitchell}. 

\acknowledgments  This work was supported by Research Innovation Award R10815 of the Research Corporation and by the Packard Foundation.


\begin{thebibliography}{41}
\expandafter\ifx\csname natexlab\endcsname\relax\def\natexlab#1{#1}\fi
\expandafter\ifx\csname bibnamefont\endcsname\relax
  \def\bibnamefont#1{#1}\fi
\expandafter\ifx\csname bibfnamefont\endcsname\relax
  \def\bibfnamefont#1{#1}\fi
\expandafter\ifx\csname citenamefont\endcsname\relax
  \def\citenamefont#1{#1}\fi
\expandafter\ifx\csname url\endcsname\relax
  \def\url#1{\texttt{#1}}\fi
\expandafter\ifx\csname urlprefix\endcsname\relax\def\urlprefix{URL }\fi
\providecommand{\bibinfo}[2]{#2}
\providecommand{\eprint}[2][]{\url{#2}}

\bibitem[{\citenamefont{{A. Mizel}}(2001)}]{Mizel01}
\bibinfo{author}{\bibnamefont{{A. Mizel, M. W. Mitchell, and M. L. Cohen}}}, \bibinfo{journal}{Phys. Rev. A}
\textbf{\bibinfo{volume}{63}}, \bibinfo{pages}{04030}2(\bibinfo{year}{R}) (2001).

\bibitem[{\citenamefont{{A. Mizel}}(2002)}]{Mizel02}
\bibinfo{author}{\bibnamefont{{A. Mizel, M. W. Mitchell, and M. L. Cohen}}}, \bibinfo{journal}{Phys. Rev. A}
\textbf{\bibinfo{volume}{65}}, \bibinfo{pages}{022315} (\bibinfo{year}{2002}).

\bibitem[{\citenamefont{M. A. Nielsen and I. L. Chuang}(1994)}]{Nielsen00}
\bibinfo{author}{\bibfnamefont{M. A. Nielsen and I. L. Chuang}},
  \emph{\bibinfo{title}{{Quantum Computation and Quantum Information}}}
  (\bibinfo{publisher}{{Cambridge Univ. Press}}, \bibinfo{address}{{Cambridge}}, \bibinfo{year}{2000}).

\bibitem[{\citenamefont{{P.W. Shor}}(1994)}]{Shor94}
\bibinfo{author}{\bibnamefont{{P.W. Shor}}}, in
  \emph{\bibinfo{booktitle}{{Proceedings of the 35th Symposium on Foundations of Computing}}} (\bibinfo{publisher}{{IEEE Computer Society Press}},
  \bibinfo{address}{{New York}}, \bibinfo{year}{1994}),  p.~\bibinfo{pages}{124}.

\bibitem[{\citenamefont{{A. Ekert and R. Jozsa}}(1996)}]{Ekert96}
\bibinfo{author}{\bibnamefont{{A. Ekert and R. Jozsa}}}, \bibinfo{journal}{Rev. Mod. Phys.}
\textbf{\bibinfo{volume}{68}}, \bibinfo{pages}{1} (\bibinfo{year}{1996}).

\bibitem[{\citenamefont{{L.K. Grover}}(1996)}]{Grover96}
\bibinfo{author}{\bibnamefont{{L.K. Grover}}}, in
  \emph{\bibinfo{booktitle}{Proceedings of the 28th Annual ACM Symposium on the Theory of Computation}} (\bibinfo{publisher}{{ACM Press}},
  \bibinfo{address}{{New York}}, \bibinfo{year}{1996}),  p.~\bibinfo{pages}{212}.

\bibitem[{\citenamefont{{L. K. Grover}}(1997)}]{Grover97}
\bibinfo{author}{\bibnamefont{{L. K. Grover}}}, \bibinfo{journal}{Phys. Rev. Lett.}
\textbf{\bibinfo{volume}{79}}, \bibinfo{pages}{325} (\bibinfo{year}{1997}).

\bibitem[{\citenamefont{{D. Deutsch}}(1985)}]{Deutsch85}
\bibinfo{author}{\bibnamefont{{D. Deutsch}}}, \bibinfo{journal}{Proc. R. Soc. London Ser. A}
\textbf{\bibinfo{volume}{400}}, \bibinfo{pages}{97} (\bibinfo{year}{1985}).

\bibitem[{\citenamefont{{R. Jozsa}}(1991)}]{Jozsa91}
\bibinfo{author}{\bibnamefont{{R. Jozsa}}}, \bibinfo{journal}{ Proc. R. Soc. London Ser. A}
\textbf{\bibinfo{volume}{435}}, \bibinfo{pages}{563} (\bibinfo{year}{1991}).

\bibitem[{\citenamefont{{J. I. Cirac and P. Zoller}}(1995)}]{Cirac95}
\bibinfo{author}{\bibnamefont{{J. I. Cirac and P. Zoller}}}, \bibinfo{journal}{Phys. Rev. Lett.}
\textbf{\bibinfo{volume}{74}}, \bibinfo{pages}{4091} (\bibinfo{year}{1995}).

\bibitem[{\citenamefont{{C. Monroe {\it et al.}}}(1995)}]{Monroe95}
\bibinfo{author}{\bibnamefont{{C. Monroe {\it et al.}}}}, \bibinfo{journal}{Phys. Rev. Lett.}
\textbf{\bibinfo{volume}{75}}, \bibinfo{pages}{4714} (\bibinfo{year}{1995}).

\bibitem[{\citenamefont{{Q. A. Turchette, C. J. Hood, W. Lange, H. Mabuchi, and H. J. Kimble}}(1995)}]{Turchette95}
\bibinfo{author}{\bibnamefont{{Q. A. Turchette, C. J. Hood, W. Lange, H. Mabuchi, and H. J. Kimble}}}, \bibinfo{journal}{Phys. Rev. Lett.}
\textbf{\bibinfo{volume}{75}}, \bibinfo{pages}{4710} (\bibinfo{year}{1995}).

\bibitem[{\citenamefont{{N. Gershenfeld and I. L. Chuang}}(1997)}]{Gershenfeld97}
\bibinfo{author}{\bibnamefont{{N. Gershenfeld and I. L. Chuang}}}, \bibinfo{journal}{Science}
\textbf{\bibinfo{volume}{275}}, \bibinfo{pages}{350} (\bibinfo{year}{1997}).

\bibitem[{\citenamefont{{I. L. Chuang}}(1998)}]{Chuang98}
\bibinfo{author}{\bibnamefont{{I. L. Chuang, N. Gershenfeld, M. Kubinec}}}, \bibinfo{journal}{Phys. Rev. Lett.}
\textbf{\bibinfo{volume}{80}}, \bibinfo{pages}{3408} (\bibinfo{year}{1998}).

\bibitem[{\citenamefont{{D. Loss and D. P. DiVincenzo}}(1998)}]{Loss98}
\bibinfo{author}{\bibnamefont{{D. Loss and D. P. DiVincenzo}}}, \bibinfo{journal}{Phys. Rev. A}
\textbf{\bibinfo{volume}{57}}, \bibinfo{pages}{120} (\bibinfo{year}{1998}).

\bibitem[{\citenamefont{{G. Burkhard}}(1999)}]{Burkhard99}
\bibinfo{author}{\bibnamefont{{G. Burkhard, D. Loss, D. P. DiVincenzo}}}, \bibinfo{journal}{Phys. Rev. B}
\textbf{\bibinfo{volume}{59}}, \bibinfo{pages}{2070} (\bibinfo{year}{1999}).

\bibitem[{\citenamefont{{A. Shnirman}}(1997)}]{Shnirman97}
\bibinfo{author}{\bibnamefont{{A. Shnirman, G. Schon, Z. Hermon}}}, \bibinfo{journal}{Phys. Rev. Lett.}
\textbf{\bibinfo{volume}{79}}, \bibinfo{pages}{2371} (\bibinfo{year}{1997}).

\bibitem[{\citenamefont{{Y. Makhlin}}(1999)}]{Makhlin99}
\bibinfo{author}{\bibnamefont{{Y. Makhlin, G. Schon, A. Shnirman}}}, \bibinfo{journal}{Nature}
\textbf{\bibinfo{volume}{398}}, \bibinfo{pages}{305} (\bibinfo{year}{1999}).

\bibitem[{\citenamefont{{D. V. Averin}}(1998)}]{Averin98}
\bibinfo{author}{\bibnamefont{{D. V. Averin}}}, \bibinfo{journal}{Solid State Commun.}
\textbf{\bibinfo{volume}{105}}, \bibinfo{pages}{659} (\bibinfo{year}{1998}).

\bibitem[{\citenamefont{{Y. Nakamura}}(1999)}]{Nakamura99}
\bibinfo{author}{\bibnamefont{{Y. Nakamura, Yu. A. Pashkin, J. S. Tsai}}}, \bibinfo{journal}{Nature}
\textbf{\bibinfo{volume}{398}}, \bibinfo{pages}{786} (\bibinfo{year}{1999}).

\bibitem[{\citenamefont{{J. E. Mooij {\it et al.}}}(1999)}]{Mooij99}
\bibinfo{author}{\bibnamefont{{J. E. Mooij {\it et al.}}}}, \bibinfo{journal}{Science}
\textbf{\bibinfo{volume}{285}}, \bibinfo{pages}{1036} (\bibinfo{year}{1999}).

\bibitem[{\citenamefont{{L. B. Ioffe}}(1999)}]{Ioffe99}
\bibinfo{author}{\bibnamefont{{L. B. Ioffe, V. B. Geshkenbein, M. V. Feigel'man, A. L. Fauchere, G. Blatter}}}, \bibinfo{journal}{Nature}
\textbf{\bibinfo{volume}{398}}, \bibinfo{pages}{679} (\bibinfo{year}{1999}).

\bibitem[{\citenamefont{{D. Vion}}(2002)}]{Vion02}
\bibinfo{author}{\bibnamefont{{D. Vion, A. Aassime, A. Cottet, P. Joyez, H. Pothier, C. Urbina, D. Esteve, M. H. Devoret}}}, \bibinfo{journal}{Science}
\textbf{\bibinfo{volume}{296}}, \bibinfo{pages}{886} (\bibinfo{year}{2002}). 

\bibitem[{\citenamefont{{Y. A. Pashkin}}(2003)}]{Pashkin03}
\bibinfo{author}{\bibnamefont{{Y. A. Pashkin, T. Yamamoto, O. Astafiev, Y. Nakamura, D. V. Averin, J. S. Tsai}}}, \bibinfo{journal}{Nature}
\textbf{\bibinfo{volume}{421}}, \bibinfo{pages}{823} (\bibinfo{year}{2003}).

\bibitem[{\citenamefont{{B. Kane}}(1998)}]{Kane98}
\bibinfo{author}{\bibnamefont{{B. Kane}}}, \bibinfo{journal}{Nature}
\textbf{\bibinfo{volume}{393}}, \bibinfo{pages}{133} (\bibinfo{year}{1998}).

\bibitem[{\citenamefont{{A. Barenco et al.}}(1995)}]{Barenco95}
\bibinfo{author}{\bibnamefont{{A. Barenco et al.}}}, \bibinfo{journal}{Phys. Rev. A}
\textbf{\bibinfo{volume}{52}}, \bibinfo{pages}{3457} (\bibinfo{year}{1995}).

\bibitem[{\citenamefont{{W. K. Wootters and W. H. Zurek}}(1982)}]{Wootters82}
\bibinfo{author}{\bibnamefont{{W. K. Wootters and W. H. Zurek}}}, \bibinfo{journal}{Nature}
\textbf{\bibinfo{volume}{299}}, \bibinfo{pages}{802} (\bibinfo{year}{1982}).

\bibitem[{\citenamefont{{R. P. Feynman}}(1985)}]{Feynman85}
\bibinfo{author}{\bibnamefont{{R. P. Feynman}}}, \bibinfo{journal}{Optics News}
\textbf{\bibinfo{volume}{11:2}}, \bibinfo{pages}{11} (\bibinfo{year}{1985}).

\bibitem[{\citenamefont{{A. Peres}}(1985)}]{Peres85}
\bibinfo{author}{\bibnamefont{{A. Peres}}}, \bibinfo{journal}{Phys. Rev. A}
\textbf{\bibinfo{volume}{32}}, \bibinfo{pages}{3266} (\bibinfo{year}{1985}).

\bibitem[{\citenamefont{{C. H. Bennett et al.}}(1993)}]{Bennett93}
\bibinfo{author}{\bibnamefont{{C. H. Bennett et al.}}}, \bibinfo{journal}{Phys. Rev. Lett.}
\textbf{\bibinfo{volume}{70}}, \bibinfo{pages}{1895} (\bibinfo{year}{1993}).

\bibitem[{\citenamefont{{M. A. Nielsen and I. L. Chuang}}(1997)}]{Nielsen97}
\bibinfo{author}{\bibnamefont{{M. A. Nielsen and I. L. Chuang}}}, \bibinfo{journal}{Phys. Rev. Lett.}
\textbf{\bibinfo{volume}{79}}, \bibinfo{pages}{321} (\bibinfo{year}{1997}).

\bibitem[{\citenamefont{{D. Gottesman and I. L. Chuang}}(1999)}]{Gottesman99}
\bibinfo{author}{\bibnamefont{{D. Gottesman and I. L. Chuang}}}, \bibinfo{journal}{Nature}
\textbf{\bibinfo{volume}{402}}, \bibinfo{pages}{390} (\bibinfo{year}{1999}).

\bibitem[{\citenamefont{{G. Vidal}}(2002)}]{Vidal02}
\bibinfo{author}{\bibnamefont{{G. Vidal, L. Masanes, and J. I. Cirac}}}, \bibinfo{journal}{Phys. Rev. Lett.}
\textbf{\bibinfo{volume}{88}}, \bibinfo{pages}{047905} (\bibinfo{year}{2002}).

\bibitem[{\citenamefont{{greaterthan}}(2003)}]{greaterthan}
\bibinfo{author}{\bibnamefont{{Since our controlled-NOT gate (\ref{CNOT}) ensures that the target bit is always at a lower row than the control bit, the probabilities are actually greater than our estimates.}}}

\bibitem[{\citenamefont{{Mitchell}}(2003)}]{Mitchell}
\bibinfo{author}{\bibnamefont{{This idea is due to M. W. Mitchell, who refers to it as the ``squeegee'' approach because of the way that potential pushes the ground state down the quantum dot array.}}}

\end{thebibliography}
\end{document}